\documentclass[preprint,12pt]{elsarticle}

\usepackage{booktabs}
\usepackage{multirow}
\usepackage{amssymb}
\usepackage{amsmath}

\journal{arXiv}

\begin{document}

\begin{frontmatter}

\title{Practice Support for Violin Bowing by Measuring Bow Pressure and Position}

\author[label1]{Yurina Mizuho} 
\author[label1]{Yuta Sugiura}
\affiliation[label1]{organization={Keio University},
            addressline={3-14-1 Hiyoshi, Kohoku-ku}, 
            city={Yokohama},
            postcode={223-8522}, 
            state={Kanagawa},
            country={Japan}}

\begin{abstract}
The violin is one of the most popular musical instruments. Various parameters of bowing motion, such as pressure, position, and speed, are crucial for producing a beautiful tone. However, mastering them is challenging and requires extensive practice. In this study, we aimed to support practice of bowing, focusing on bow pressure. First, we compared the bowing movements, specifically bow pressure, bow position, and bow speed, of eight experienced players with those of eight beginners. Next, we developed and evaluated a visual feedback system that displays bow pressure to support practice. We taught the identified differences to 14 beginners, dividing them into two groups: one practiced with an explanation, and the other with both an explanation and a feedback system. These two experiments found that clarifying the characteristics unique to experienced players can support practice.
\end{abstract}


\end{frontmatter}



\section{Introduction}
Bowed string instruments, such as the violin, viola, and cello,  produce sound by pressing the strings with the left hand and bowing the strings with the right hand. With these instruments, the bowing action has a significant impact on tonal quality~\cite{Cremer,Fischer,Galamian}. The tone results from a delicate balance of multiple bowing parameters, including bow speed, the distance between the bow and bridge, bow pressure, bow angle, and bow trajectory. Many players adjust their playing movements based on their own experience and intuition. Therefore, achieving a wide variety of tones on the violin requires extensive practice time and experience. Also, unlike other stringed instruments such as piano and guitar, bowing is an unusual movement, making it difficult to produce a beautiful sound immediately, which is a very high hurdle for beginners.

Much research has analyzed the movements of skilled players to support effective instruction and practice for learning special physical skills such as sports~\cite{Chang,Iino,Reid} and playing musical instruments~\cite{Furuya,cello}. Some studies have focused on skilled violin players; Volpe et al.~\cite{Volpe} have recorded the movements of professional violin players using motion capture. Many studies have also compared experienced and novice players to extract features in common with skilled players. Kinoshita et al.~\cite{Kinoshita} measured the force of pressing the strings with fingers using a force sensor installed on the fingerboard and compared the results. The studies which focused on player movement measured it using motion capture. They found differences in upper body movement~\cite{DAmato}, shoulder joint angle~\cite{Konczak}, bow trajectory~\cite{musicjacket}, bow angle~\cite{Blanco}, and so on. Some of these studies applied the results to a practice support system~\cite{musicjacket,Blanco} and verified the effectiveness of practice.

For the violin bowing, the balance of three parameters mainly determines tonal qualify: the bow speed, the distance between the bow and the bridge, and bow pressure on the string~\cite{Fischer}. As bow pressure is particularly less visually apparent than other parameters, they require auditory evaluation, which can be challenging for beginners and those practicing independently whose ears are not well trained. However, no studies have analyzed the characteristics of skilled players concerning bow pressure. Bow pressure is often measured using strain gauges~\cite{askenfelt1,askenfelt2,Young} and optical sensors~\cite{Pardue}. However, these studies have aimed to improve the accuracy and simplicity of bow pressure measurement and have not analyzed bow pressure during performance.

\begin{figure}[tb]
\centering
\includegraphics[width=0.6\columnwidth]{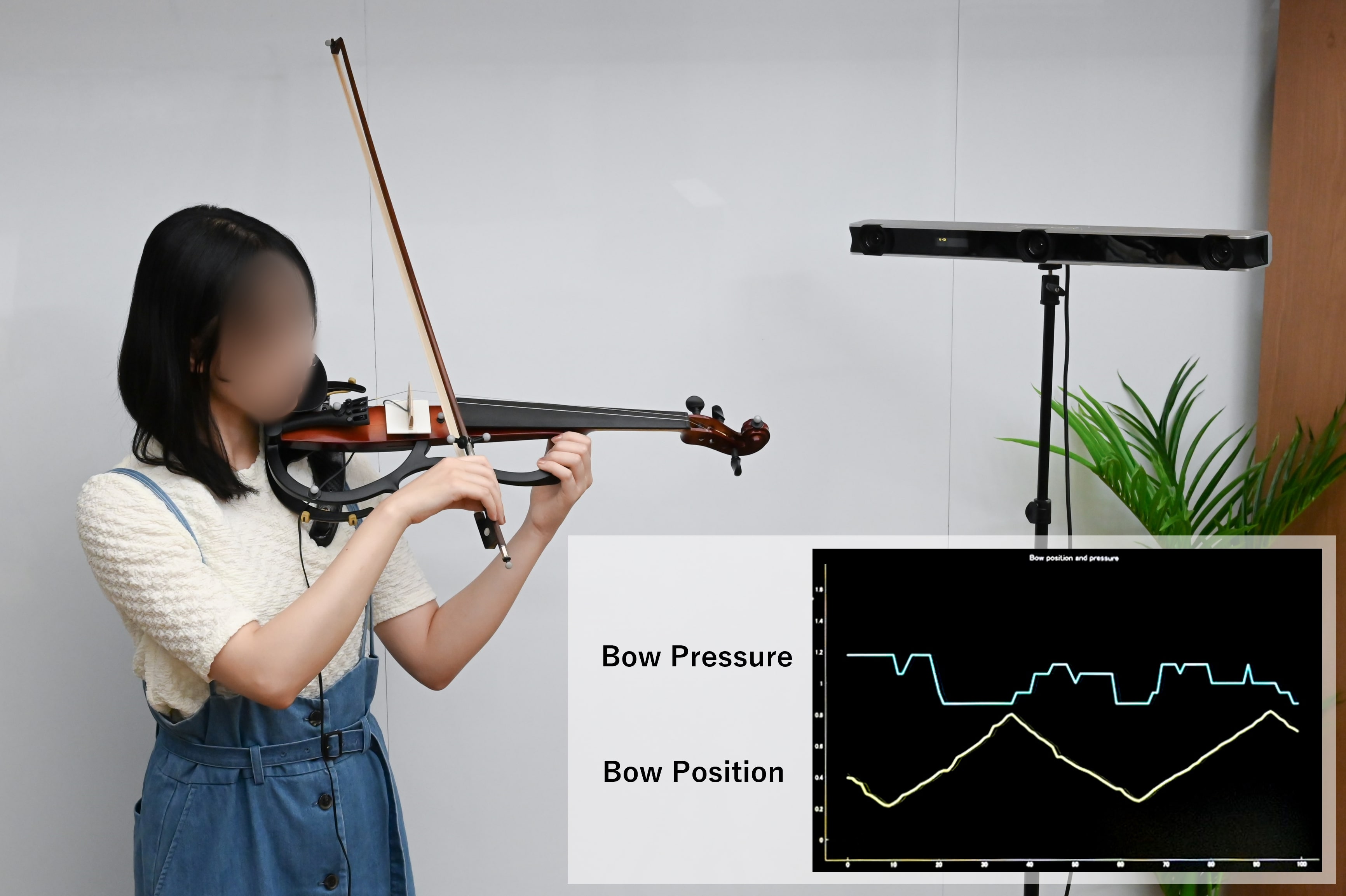}
\caption{Bow pressure and position measurement}
\label{fig:thumbnail}
\end{figure}

Therefore, in this study, we focused on bow pressure among bowing motions. First, we compared it among experienced players and beginners (Figure~\ref{fig:thumbnail}). We measured bow pressure using a force sensor and bow position using motion capture. The results of the difference analysis showed that the experienced players always maintained a certain level of bow pressure and kept the bow pressure even at the tip of the bow.

Second, we verified whether explaining the differences to beginners and practicing with a system to support improving the differences would affect bowing practice. The results showed that the bow pressure increased significantly in the post-teaching measurement, and the users tended to try to reduce the difference in bow pressure at each bow position. We found that teaching the characteristics common to experienced players had some effect on practice.

\section{Related works}
\label{sec:relatedworks}
\subsection{Performance Analysis for Skills Training}
\label{sec:skilltraining}
Several studies analyze the movements of skilled players to develop effective teaching and training methods for acquiring special physical skills. Measuring and comparing the movements of players of various skill levels can quantitatively clarify characteristics specific to skilled players. The results can increase the efficiency and persuasiveness of instruction and serve as an indicator of achievement during self-practice.

For example, in the field of sports, Reid et al.~\cite{Reid} used motion capture to capture the movements of skilled tennis players and analyzed differences in the movements of racket, ball, and hand for different types of shots; Iino et al.~\cite{Iino} measured the movements of advanced and intermediate table tennis players and found that advanced players were able to accelerate their rackets more quickly. Chang et al.~\cite{Chang} measured the movements of novice, intermediate, and expert dancers and found differences in the amplitude and speed of joint movements.

In the field of instrumental performance, there have also been many studies comparing experienced and novice players: Furuya et al.~\cite{Furuya} showed that the movement of the upper limb joints during key striking differs between skilled and novice piano players. Verrel et al.~\cite{cello} also found that skilled cello players consistently maintained a bow-to-string angle close to 90 degrees and exhibited smaller variations in joint angles of the elbow and wrist.

\subsection{Analysis of Violin Players' Movements}
\label{sec:violintraining}
In violin as well, studies have focused on the playing motion of skilled players for practice support and recording;  Volpe et al.~\cite{Volpe} captured and recorded the sounds and movements of professional violinists using motion capture, video, sound, and Kinect, and published their data for practice support and educational purposes. There are also many studies comparing experienced and novice players to identify the characteristics of experienced players. Kinoshita et al.~\cite{Kinoshita} used force sensors embedded in the fingerboard to compare the press forces of experienced and novice players; D'Amato et al.~\cite{DAmato} discriminated between experienced and novice players based on their playing movements measured by motion capture and showed that the features of upper body movements were particularly important. Konczak et al.~\cite{Konczak} used motion capture to measure bow, shoulder, and elbow angles during violin playing in experienced and novice players and found that the amplitude of shoulder joint angles of experienced players was smaller.

Some studies applied the results of such comparisons to the development of practice support systems. For example, Blanco et al.~\cite{Blanco} found differences in sound stability, pitch stability, and bow trajectory between experienced and novice players and developed a smartphone app that displays sound quality and bow angle. Van der Linden et al.~\cite{musicjacket} found differences between experienced and novice players in bow trajectory using motion capture; and they developed a system to provide tactile feedback when a player's bow deviates from the desired trajectory. These studies have also tested the effectiveness of the system in practice.

\begin{table}
    \centering
    \resizebox{\textwidth}{!}{%
    \begin{tabular}{r|ccccc}
         \toprule
         Related Works & Bowing & Measurement & Comparison & Application \\
         \hline
         Konczak et al.~\cite{Konczak} & Angle & Motion capture & $\circ$ & $\times$ \\
         Blanco et al.~\cite{Blanco}& Angle & Motion capture & $\circ$  & Application\\
         Music Jacket~\cite{musicjacket}& Trajectory & Motion capture & $\circ$ & Vibration \\
         \hline
         Askenfelt~\cite{askenfelt1}& Pressure & Strain gauge & $\times$ & $\times$ \\
         Young~\cite{Young}& Pressure & Strain gauge & $\times$ & $\times$\\
         Pardue et al.~\cite{Pardue}& Pressure & Optical sensor & $\times$ & $\times$\\
         Tanjo et al.~\cite{tanjo} & Pressure & Strain gauge & $\times$ & Graph \\
         \hline
         \hline
         \multirow{2}{*}{Our study}  & Pressure & Force sensor & $\circ$ & Explanation \\
          & Position & Motion capture &  & + System \\
         \bottomrule
    \end{tabular}}
    \caption{Comparison of our study with related works that measure bowing}
    \label{tab:related}
\end{table}

Table~\ref{tab:related} summarizes studies on the measurement and analysis of violin bowing and the position of our study. There has been no detailed analysis of bow pressure in skilled players, although bow pressure is one of the particularly important elements in violin bowing motion~\cite{Cremer,Fischer,Galamian}. Tanjo et al.~\cite{tanjo} studied a practice support focusing on bow pressure. They used motion capture and strain gauges to measure the player's arm joint angles and bow pressure and constructed a system that displays graphs of these data and a video of the player. However, this study only displayed the measured bow pressure on the screen and did not analyze the bow pressure of skilled players in detail or verify the effect of practice. 

Therefore, we compared and analyzed the bowing motions of experienced players and beginners, focusing especially on bow pressure. Then, we verified whether teaching the results led to practice support.

\subsection{Bowing Measurement for Violins}
\label{sec:related method}
Bowing action has a significant impact on tonal quality. Cremer~\cite{Cremer} executed physical analyses of violins to explore how different playing parameters influence timbre. For the violin, the balance of three parameters mainly determines tonal qualify: the bow speed, the distance between the bow and the bridge, and bow pressure on the string~\cite{Cremer,Fischer,Galamian}. Many studies have been conducted to measure bowing motion quantitatively for performance recording, analysis, and practice support. They often use motion capture to measure arm and body movements during a performance. This method is applied to record~\cite{Volpe}, analyze performance movements~\cite{DAmato,Konczak}, and feedback systems~\cite{musicjacket}, as described in section~\ref{sec:violintraining}.

Bow position, speed, and angle can also be measured with motion capture; the system of Blanco et al.~\cite{Blanco} used motion capture to measure bow angles and presented them to a smartphone app. Schoonderwaldt and Demoucron~\cite{mocap} used motion capture to measure many factors such as bow position, bow speed, bow angle, and distance between bow and bridge. Other methods use resistance wires~\cite{askenfelt1} or optical sensors~\cite{Provenzale} to measure bow position. 

Related studies measuring bow pressure often use strain gauges~\cite{askenfelt1,Young,tanjo,Demoucron} and optical sensors~\cite{Pardue}. These methods estimate bow pressure from the bow deformation. These studies have focused on the simplicity and accuracy of the measurement methods and have not analyzed the bow pressure of performers in detail.

In our study, we measured the bow pressure, bow position, and bow speed of experienced players and beginners. We used a force sensor to measure bow pressure; Pardue et al.~\cite{Pardue} measured bow pressure with force sensors as a true value to build an estimation model of bow pressure. We measured bow position and speed using motion capture which is the most commonly used.

\section{Measurement Methods}
\subsection{Bow Pressure}
\begin{figure}[tb]
\centering
\includegraphics[width=0.5\columnwidth]{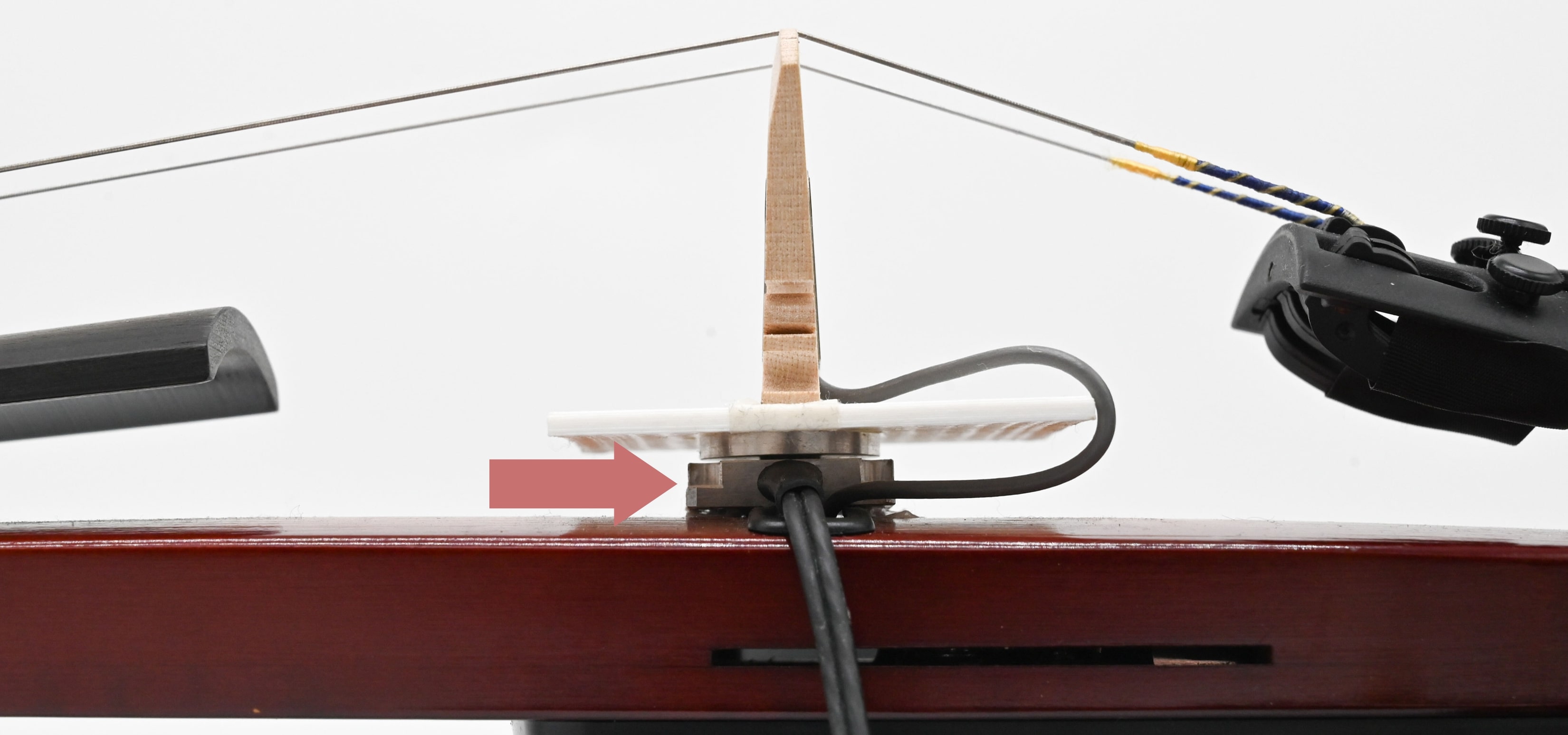}
\caption{Load cell installed under the bridge}
\label{fig:loadcell}
\end{figure}

We used a Yamaha Silent Violin (SV250) and a Yamaha bow (CBB101). We measured bow pressure using a load cell (TEC Gihan Co., USL06-H5-50N) installed under the bridge, as shown in Figure~\ref{fig:loadcell}, focusing on the z-axis force. We placed a 5 $cm$ $\times$ 5 $cm$ white plate on the load cell to facilitate the fixation of the bridge in a position perpendicular to the body of the violin. The voltage signal from the load cell was amplified using an amplifier (TEC Gihan Co., DSA-03A) and sent to a microcontroller (Arduino Pro Mini, 3.3 $V$). The microcontroller then converted the analog signal into sensor readings and calculated the force in Newtons by applying the load cell's specific calibration factor. We connected the microcontroller to a computer via USB and obtained the forces through serial communication using Python.

\subsection{Bow Position and Bow Speed}

\begin{figure}[tb]
\centering
\includegraphics[width=0.8\columnwidth]{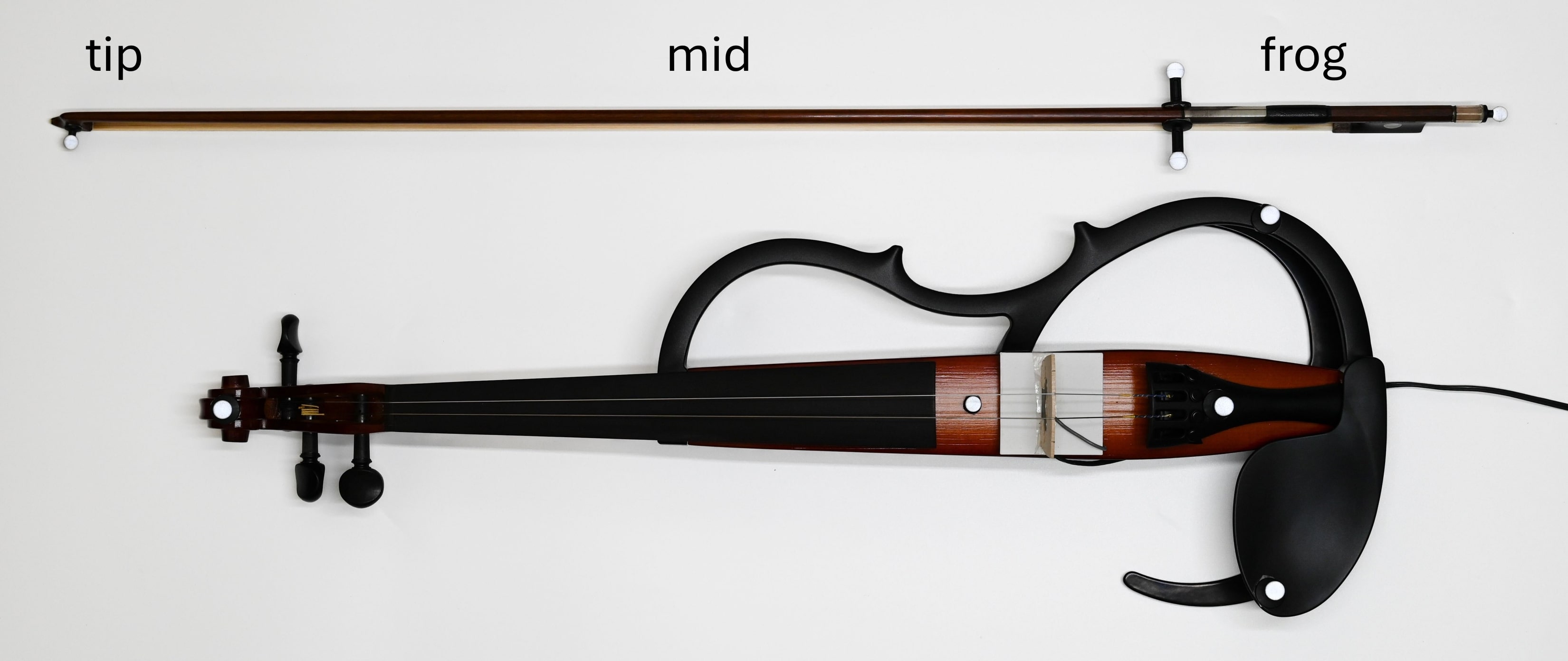}
\caption{Position of optical markers attached to the violin}
\label{fig:marker}
\end{figure}

We used an optical motion capture (Optitrack V120: Trio) to measure the bow position. We attached five markers on the violin body and five on the bow at the positions shown in Figure~\ref{fig:marker}. The arrangement was based on the study by Schoonderwaldt et al.~\cite{mocap}. We used the software Motive to obtain the coordinates of the markers and streamed the coordinates in real-time via socket communication using Python. The bow position was calculated from the coordinates of the markers. 

The bow position is where the bow hairs make contact with the string along the length of the bow. The bow position was numerically expressed, with the frog (near the player’s hand) as 0 and the tip (farthest from the player’s hand) as 1. Bow speed was calculated from the change in bow position over time using the timestamps of the acquired data.

\section{Comparison of Bowing Among Experts and Beginners}
\label{sec:begexp}
We conducted the first experiment to compare the characteristics of bowing actions between experienced players and beginners. We focused on the bow pressure, bow position, and bow speed during performance. This experiment was approved by the Ethics Committee of Keio University (Approval number: 2024-087).

\subsection{Participants}
We recruited 16 participants: eight experienced violin players (three males, five females, average age 23.4 $\pm$ 1.32 years) and eight beginners (four males, four females, average age 23.6 $\pm$ 1.93 years). The experienced players had an average of 11.9 $\pm$ 4.51 years of violin experience. Among these players, in terms of lesson experience, two had more than 15 years, three had 10 to 12 years, and three had 6 to 9 years. Among the beginners, five participants had experience with other instruments, but none had experience with bowed string instruments.

\subsection{Experimental Procedure}
The experimental procedure was as follows. First, we gave beginners a brief explanation of how to hold the violin and the fundamental movements of the bow and arm. Next, we adjusted the position, height, and angle of the motion capture camera according to the participant’s height. Then, we held a practice session to check the camera setup. Following this, the main session was repeated three times.

In each session, we first recorded the initial value of the load cell with the bow not touching the string. This initial value corresponded to the pressure on the bridge when the strings were tuned, and this value was subtracted from the actual measured data to record the bow pressure. Next, we set a metronome to a tempo of 75 bpm. After giving a signal, we asked the participants to perform a down-bow (from the frog to the tip) over four counts, followed by an up-bow (from the tip to the frog) over the next four counts and to continue this back-and-forth motion until giving the end signal. The participants played the open A string. We asked them to move the bow back and forth within the marked sections at the frog and tip. We acquired the data for bow pressure and position at 60 fps, with 1,500 frames captured per session.

\subsection{Results and Discussion}
\begin{figure}[tb]
\centering
\includegraphics[width=0.8\columnwidth]{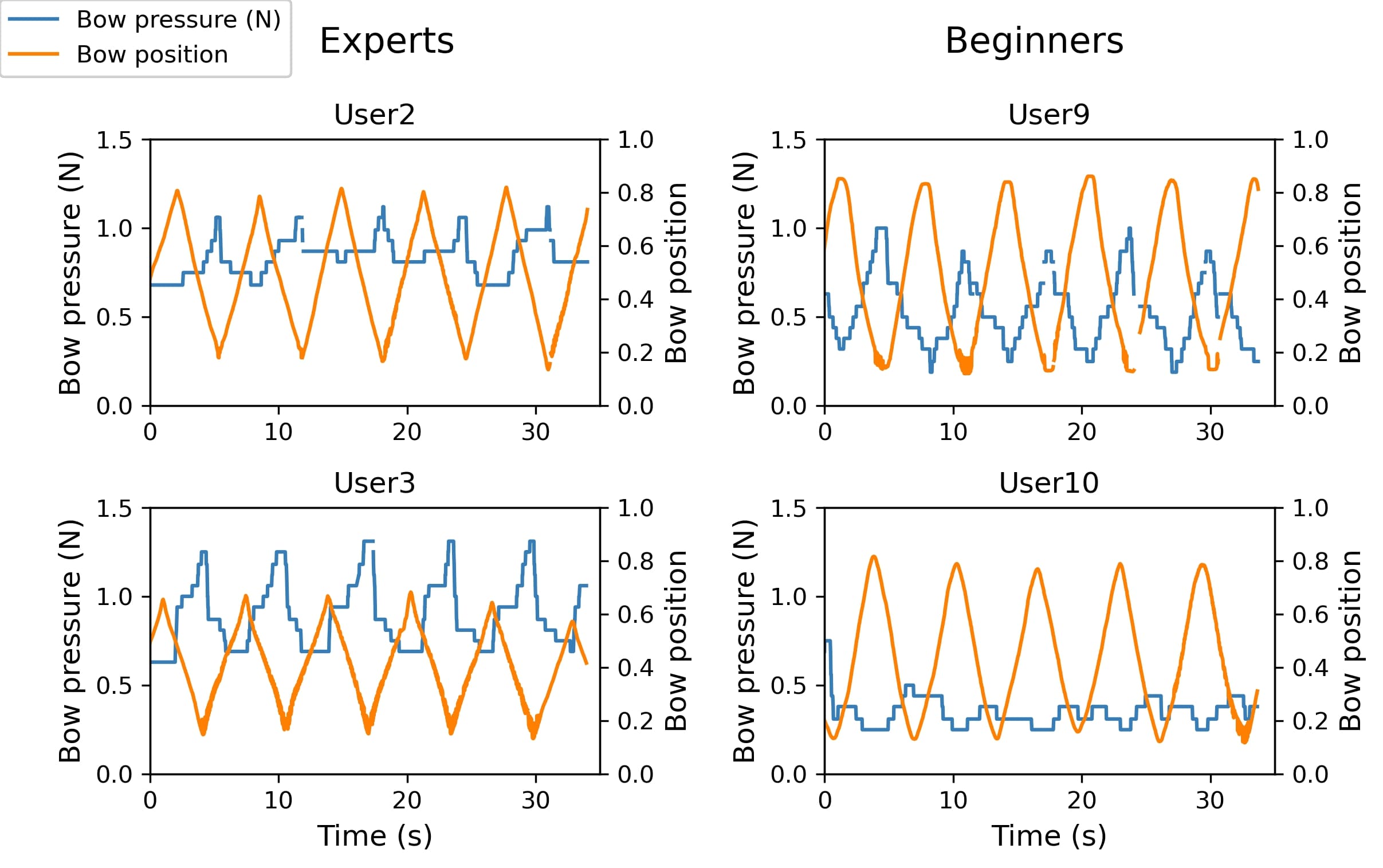}
\caption{Time series data of bow pressure and position}
\label{fig:timeseries}
\end{figure}

Figure~\ref{fig:timeseries} shows some examples of the time-series data for one trial of each participant’s main session. The horizontal axis of each graph represents time, the left vertical axis represents bow pressure, and the right vertical axis represents bow position, displayed as blue and orange graphs, respectively. The data for the experienced players are shown on the left, and the data for the beginners are presented on the right. The missing data in the graph indicates that the camera failed to capture the bow movement.

\subsubsection{Bow Pressure}
We tested the maximum value, minimum value, and range of bow pressure using the Brunner-Munzel test at a significance level of 5\%. Although there were no significant differences in bow pressure values between the experienced players and beginners, the range was larger for beginners, and both the maximum and minimum values (p = 0.059) were smaller for beginners. Seven of the eight experienced players had a minimum bow pressure of 0.5 $N$ or more. Generally, the weight of a bow is around 60 $g$, which translates to approximately 0.59 $N$ in Newton units. In this study, the experienced players usually kept bow pressure higher than or as high as the bow weight.

\begin{figure}[tb]
\centering
\includegraphics[width=0.75\columnwidth]{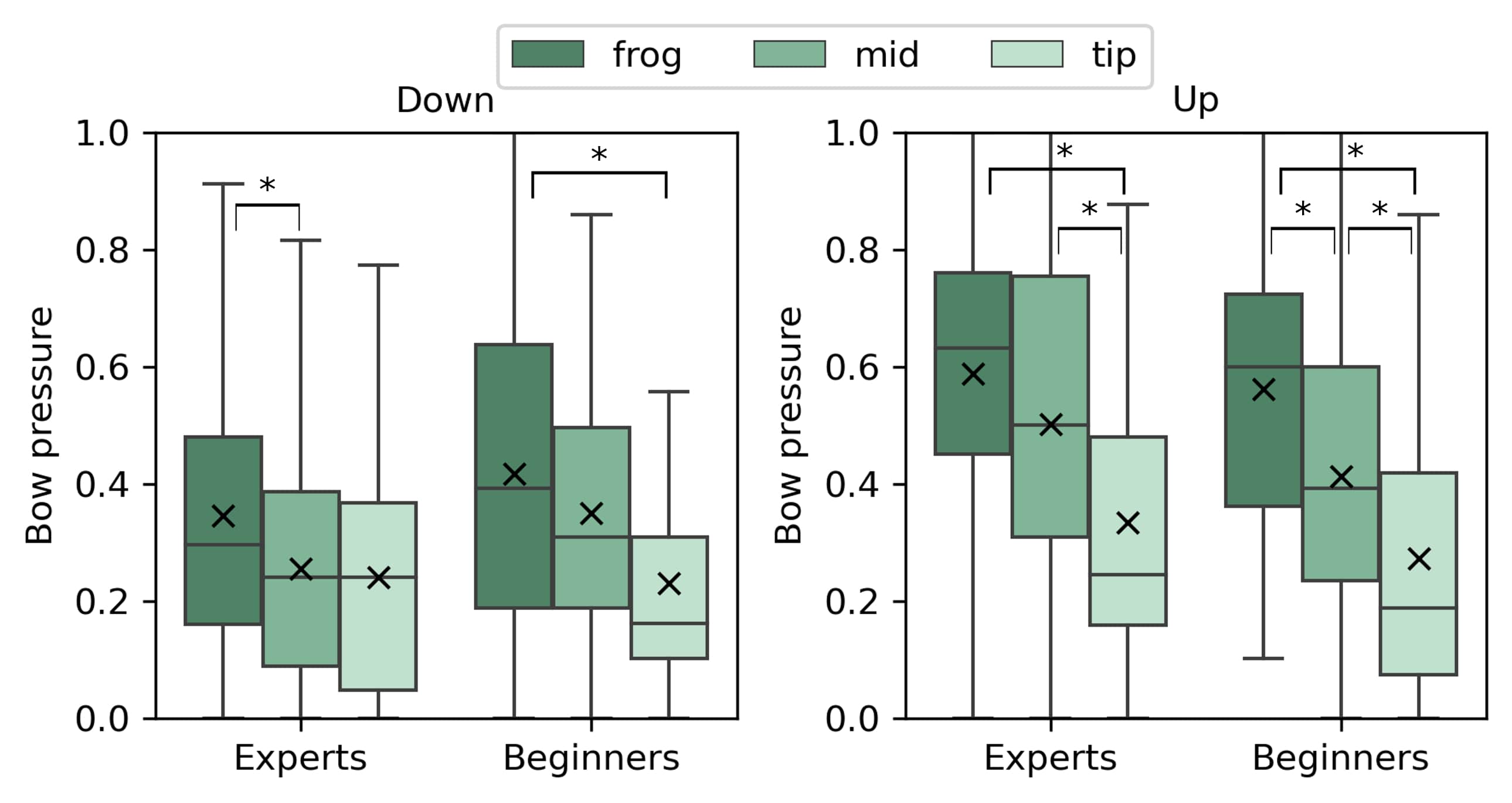}
\caption{Relationship between bow position and pressure (*: p$<$0.05)}
\label{fig:pos_pre}
\end{figure}

Next, we examined the relationship between bow position and bow pressure. Excluding the specified range of bow position, we divided the entire length of the bow into three sections: the frog, middle, and tip. Figure~\ref{fig:pos_pre} shows the bow pressure during down-bows and up-bows for experienced players and beginners. Bow pressure was normalized based on each participant’s maximum and minimum values. The Wilcoxon signed-rank test was conducted at a significance level of 5\%, with the Bonferroni correction applied for multiple comparisons of bow pressures between the three sections. During the down-bows, experienced players had significantly lower bow pressure in the middle section compared to the frog but no significant differences between the middle and tip or the frog and tip. These results suggested that the experienced players applied the necessary force at the bow frog and tried to keep consistent bow pressure, even at the tip during down-bows. We did not observe the same trend in the up-bows of the experienced players, but there was no significant difference between the frog and middle sections compared to the beginners. These results indicated that the experienced players tried to maintain consistent bow pressure from the tip toward the frog once the bow stroke started. On the other hand, beginners had significantly lower bow pressure at the tip compared to the frog, regardless of the direction of the bow stroke. As the tip is farther from the hand, the pressure could have weakened at the tip compared to that of the frog.

\begin{figure}[tb]
\centering
\includegraphics[width=0.45\columnwidth]{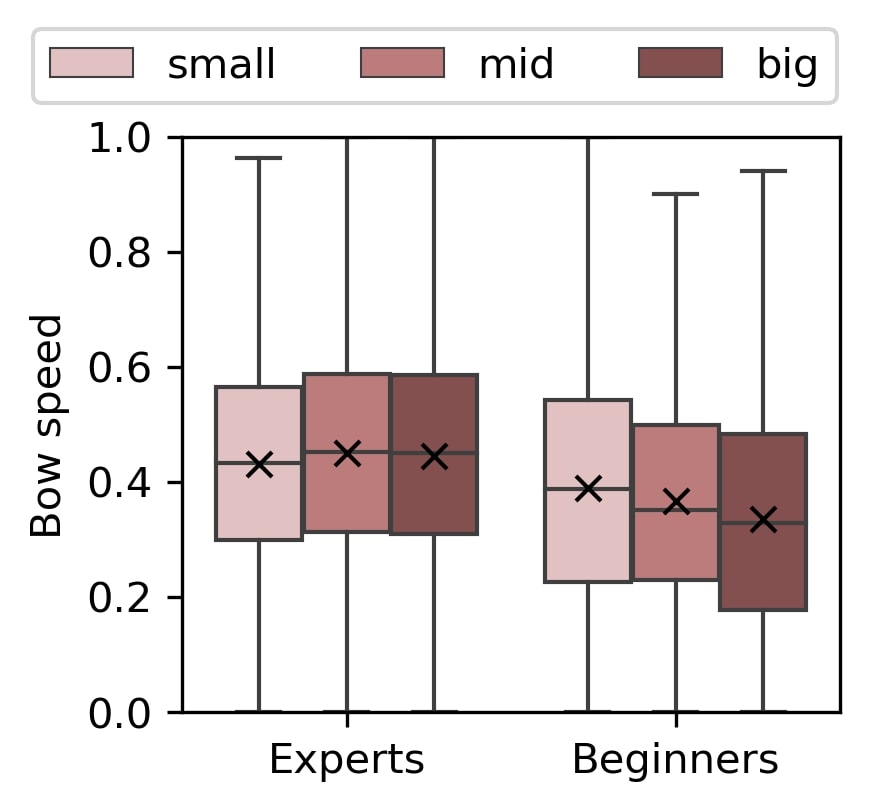}
\caption{Relationship between bow pressure and speed}
\label{fig:pr_sp}
\end{figure}

Next, we examined the relationship between bow pressure and bow speed. We normalized bow pressure and speed for each participant. We divided bow pressure into three stages based on the magnitude. We conducted the Wilcoxon signed-rank test at a significance level of 5\%, with Bonferroni correction applied for multiple comparisons of bow speeds between the three stages. As shown in Figure~\ref{fig:pr_sp}, we observed no significant differences in bow speeds between the three stages of bow pressure for either the experienced players or the beginners. However, the beginners tended to have slower bow speeds when bow pressure was high, whereas the experienced players tended to have slower bow speeds when bow pressure was low. According to Blanco et al.~\cite{Blanco}, violinists move the bow faster when applying greater force on the string to prevent the sound from becoming rough and scratchy. Conversely, when using less force, they slow down the bow movement to avoid producing a weak, airy sound lacking core resonance. In this experiment, the experienced players adjusted the relationship between bow pressure and bow speed to maintain sound quality.

\subsubsection{Bow Position}
\begin{figure}[tb]
\centering
\includegraphics[width=0.75\columnwidth]{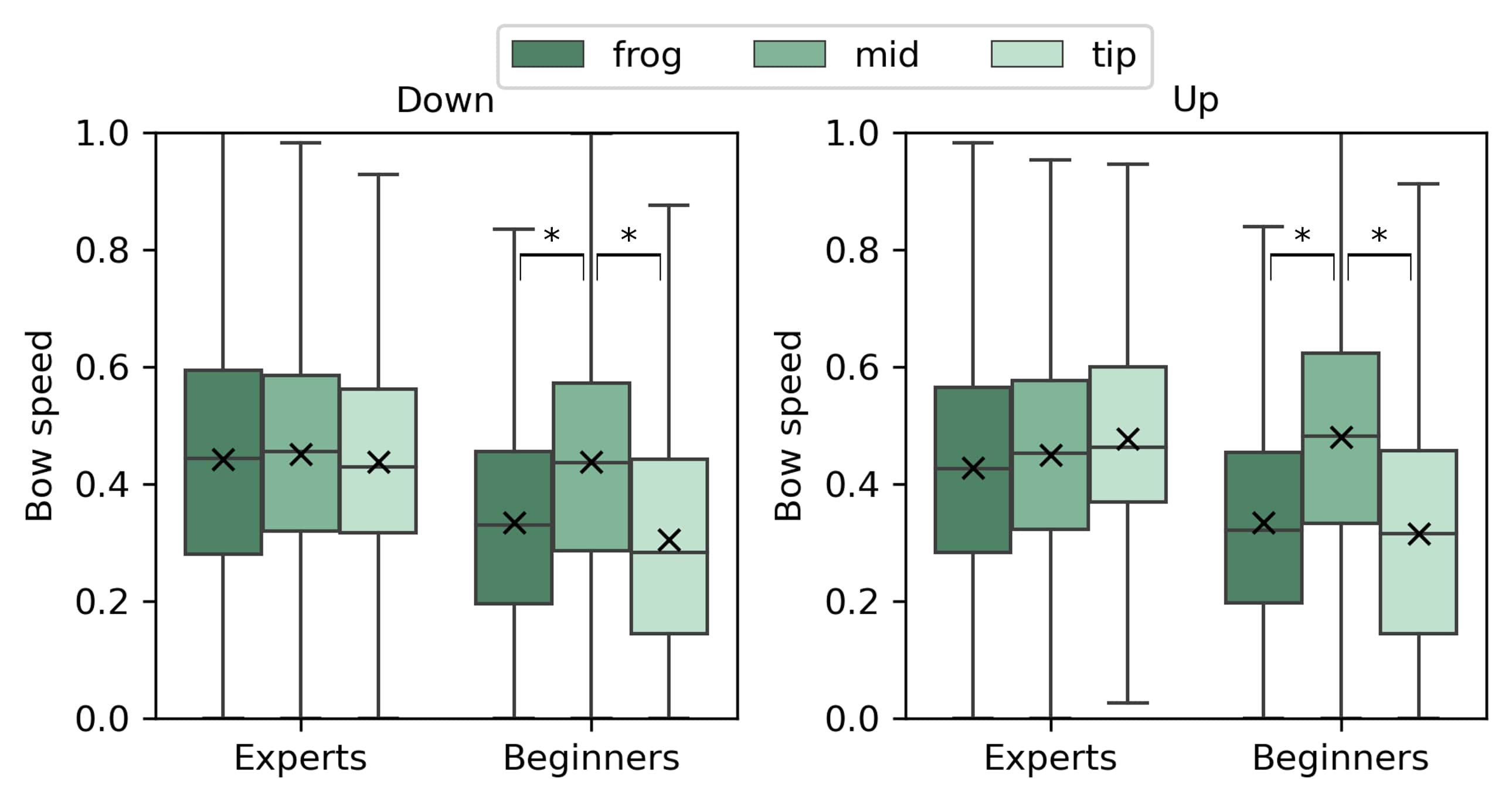}
\caption{Relationship between bow position and speed (*: p$<$0.05)}
\label{fig:pos_sp}
\end{figure}

We also examined the relationship between bow position and bow speed. Figure~\ref{fig:pos_sp} shows the bow speed during down-bows and up-bows for experienced players and beginners. Bow speed was normalized based on each participant’s maximum and minimum values. We conducted the Wilcoxon signed-rank test at a significance level of 5\%, with Bonferroni correction applied for multiple comparisons of bow speeds between three sections. Regardless of the direction of the bow stroke, we observed no significant differences in bow speed among the three sections for the experienced players. However, the beginners exhibited significant differences in bow speed between the frog and middle sections, as well as between the middle and tip sections. The beginners had faster bow speeds in the middle section and slower speeds at the frog and tip. This could be due to the difficulty in managing the pace of bowing to match the tempo.

Compared to the beginners’ graphs shown in Figure~\ref{fig:timeseries}, where the peaks at the tip and valleys at the frog indicate the direction change of the bow, the experienced players’ graphs had sharper peaks and valleys. This suggested that the beginners took longer to change their stroke direction. To verify this, we detected peaks from the time-series data of the bow position and calculated the curvature at those points. We tested the curvatures using the Brunner-Munzel test at a significance level of 5\%. The curvature was significantly larger at both the tip and the frog for experienced players. This result indicates that the experienced players could transition the direction of the bow more smoothly and quickly.


\subsubsection{Discussion}
In the bow position measurement, the beginners were unstable in the bow angle and playing posture, which caused a lot of occlusion and the camera failed to capture the bow movement. In the bow pressure measurement, the resolution of the load cell was about 0.06 $N$. Since some participants showed little change in bow pressure, it is necessary to measure it more precisely. In addition, because we installed the load cell under the bridge, the closer the bow was to the bridge, the greater the measured pressure. Therefore, a more accurate measurement of bow pressure needs a support device to allow the bow to move in the same position.

Through the experiment, we found the following four characteristics common to experienced players:

\begin{enumerate}
\renewcommand{\theenumi}{\roman{enumi}}
    \item Apply more bow pressure than the weight of the bow
    \label{p1}
    \item Keep consistent bow pressure, even at the tip
    \label{p2}
    \item Smooth change of bow direction
    \item Maintain a constant bow speed at each bow position
\end{enumerate}

These findings can be utilized, for instance, to define the acceptable range of bow pressure and speed variations, as well as to score bow direction changes.

We need to think about feedback methods. Music Jacket~\cite{musicjacket} pointed out that visual feedback requires alternating between looking at the instrument and the screen, which disperses attention and increases cognitive load. On the other hand, tactile feedback with vibration has a low cognitive load but may interfere with violin performance. Bow pressure does not necessarily have a correct value, like bow trajectory or angle. Therefore, visual feedback would be more suitable than tactile feedback because it is less forcing and does not interfere with performance. We need to select an optimal display method and devise an interface to reduce the cognitive load.

\section{Practice Support by Feedback}
\label{sec:feedback}
In Section~\ref{sec:feedback}, we investigated the effectiveness of practice by teaching beginners the differences between experienced and novice players identified in Section~\ref{sec:begexp}. We especially focused on two of the four characteristics common to experienced players related specifically to bow pressure ((\ref{p1}) Apply more bow pressure than the weight of the bow and (\ref{p2}) Keep consistent bow pressure, even at the tip). We set up two teaching methods: one group practiced after getting explanation of the differences (Group E), and the other group practiced by using a newly implemented feedback system in addition to the explanation (Group E+S). We conducted a user study to evaluate the methods. This experiment was approved by the Ethics Committee of Keio University (Approval number: 2024-149).

\subsection{Participants}
We recruited 14 violin beginners in the experiment. We divided them into two groups: a group of 7 participants (Group E) who had the explanation of differences (5 males and 2 females, average age: 23.3 $\pm$ 2.69) and a group of 7 participants (Group E+S) who used the system in addition to the explanation of differences (5 males and 2 females, average age: 23.6 $\pm$ 1.27). We decided on the group with respect to the participants' musical experience and gender. Seven participants had experience with other instruments, such as piano and guitar. One of the E+S group participants had played violin in class, but we treated him as a beginner because it was not a formal lesson and the total practice time was short. No other participants had experience with rubbed string instruments. 

\subsection{System}
\begin{figure}[tb]
\centering
\includegraphics[width=1\columnwidth]{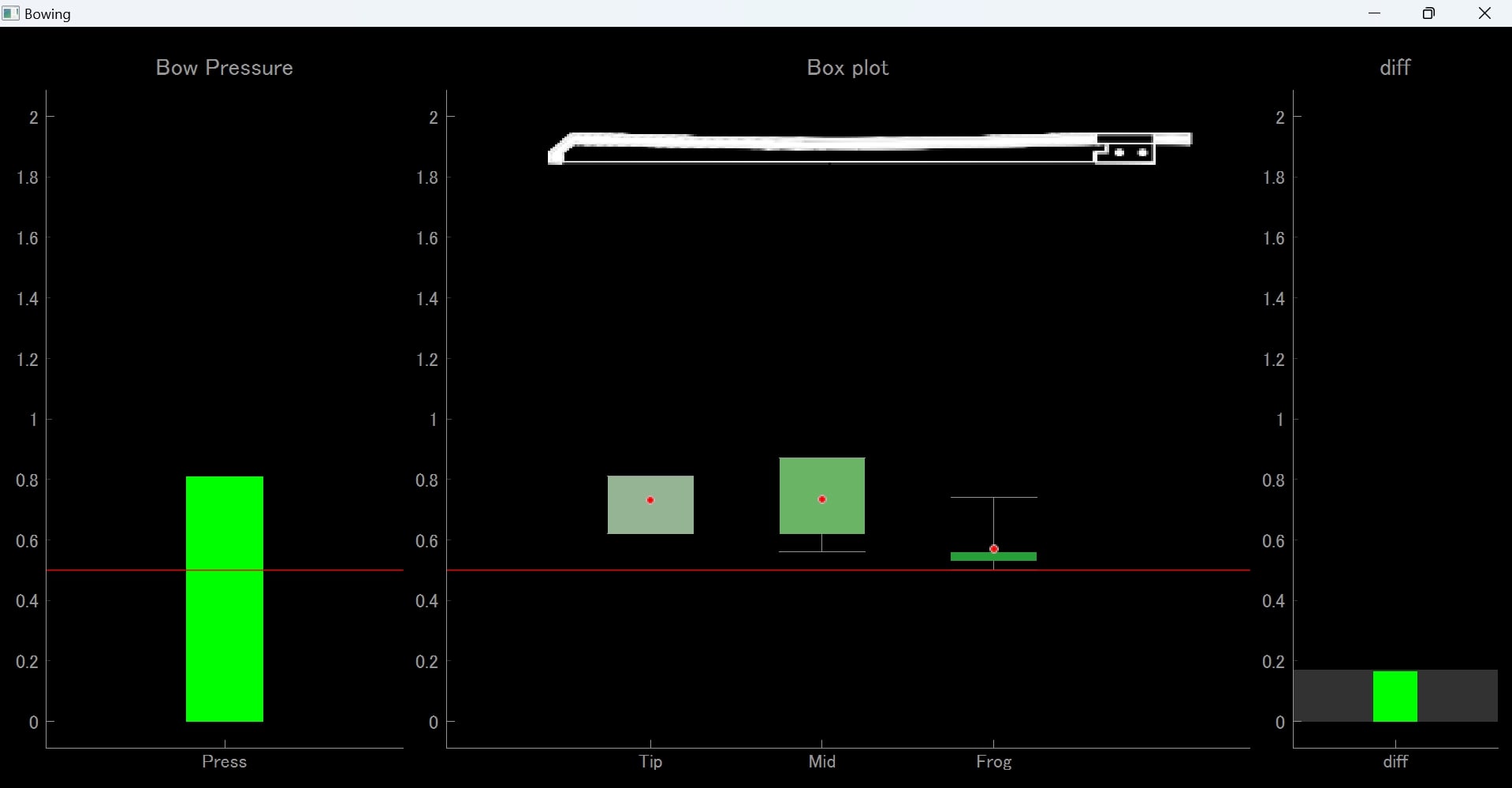}
\caption{The user interface of the visual feedback system consists of three graphs, including the bar graph showing bow pressure (left), the distribution of pressure at each bow position (center), and the difference in bow pressure at each position (right).}
\label{fig:system}
\end{figure}

As shown in Figure~\ref{fig:system}, the system consists of three graphs. The left graph is a bar graph showing the magnitude of bow pressure, updated in real-time. A red line at 0.5 $N$ shows the threshold. If the bow pressure exceeds 0.5 $N$, the bar graph is displayed in green; if the bow pressure is less than 0.5 $N$, the color changes to blue. The user should apply a bow pressure of 0.5 $N$ or more so that this graph is always green.

The central graph represents the distribution of bow pressure magnitude at each bow position, consisting of three box-and-whisker plots. This is updated when the bow is moved one round trip. The leftmost graph represents the distribution of bow pressure for one round trip at the bow tip, the middle graph at the middle of the bow, and the rightmost graph at the bow frog. The red line is also displayed at 0.5 $N$. Therefore, the users need to apply bow pressure so that the three boxes are above the red line. In addition, they should adjust the magnitude of bow pressure at each position in the next round trip so that the three boxes are as same height as each other as possible.

The right graph shows the difference between the largest mean and smallest mean of the three box plots in the center graph. This graph is updated at the same time as the center graph. A white area indicates the threshold value of 0.17 $N$, which was determined as the average of the largest difference in average bow pressure at each bow position for experienced players, as measured in the experiment in Section~\ref{sec:begexp}. If the difference exceeds the threshold, the graph color changes to red. The user should adjust the magnitude of bow pressure at each bow position so that the difference stays below the white box and the graph is always green.

\subsection{Experimental Procedure}
\begin{figure}[tb]
\centering
\includegraphics[width=\columnwidth]{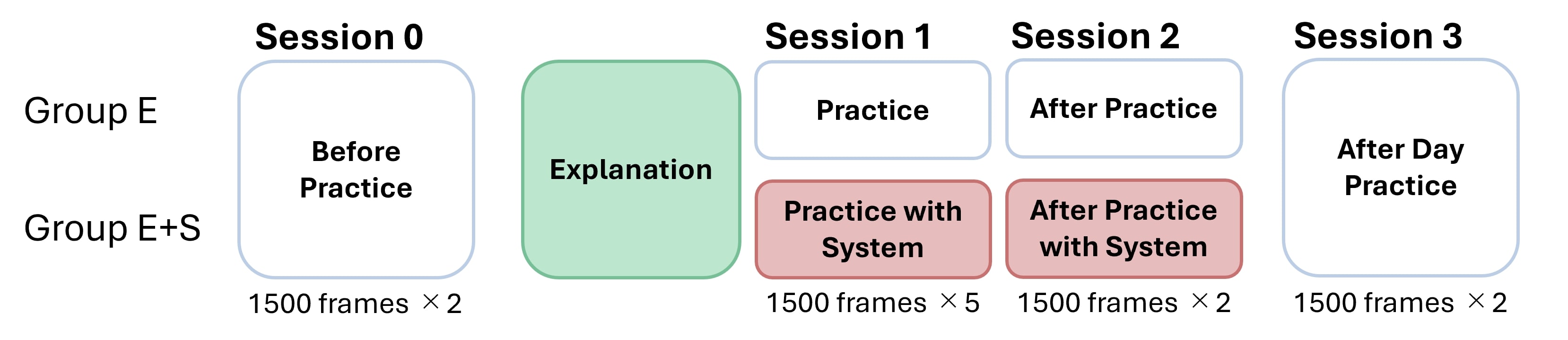}
\caption{Flow of the experiment}
\label{fig:exp}
\end{figure}

We designed the content of the experiment with reference to related studies~\cite{musicjacket,Blanco} that examined the practice effectiveness of feedback systems. First, as in the experiment in Section~\ref{sec:begexp}, we explained to participants how to hold the violin and the fundamental movements of the bow and arm. Next, we adjusted the position, height, and angle of the motion capture camera according to the participant’s height.

We conducted the experiment consisting of four sessions in total over two days (Figure~\ref{fig:exp}). First, on the first day, Session 0 (S0), we took two measurements as a baseline. Next, we explained the key points of the exercise to both groups. Specifically, we presented the experimental results in Section~\ref{sec:begexp} and instructed them to pay attention to two points: always apply bow pressure above a certain level and use the same amount of bow pressure at all bow positions, especially at the tip of the bow so that the bow pressure does not decrease. We also explained the relationship between bow pressure and sound, that too little bow pressure produces a faint sound, while too much produces a scratchy sound. Then, we explained how to use the system only for the E+S group. The participants practiced bowing during five measurements in Session 1 (S1). Only the E+S group practiced using the system. In Session 2 (S2), we took two measurements and only the E+S group used the system during the measurements. At the end of the first day, we asked the participants to answer a questionnaire. We interviewed only the E+S group about the usability of the system.

Within five days after the next day, we conducted the second day of the experiment. In Session 3 (S3), we took two measurements in both groups without the system. 

In one measurement, as in the experiment in Section~\ref{sec:begexp}, we asked the participants to continuously move the bow back and forth on the open A string to a metronome at 75 bpm. We acquired bow pressure and bow position data at 60 fps, 1,500 frames per measurement; one measurement contained data for approximately four round trips.

In the experiment in Section~\ref{sec:begexp}, beginners could not always move the bow straight and the camera sometimes failed to capture the bow movement. In addition, the accuracy of bow pressure measurement became worse because the bow sometimes shifted in the direction of the strings, significantly changing the distance between the bow and the bridge. To mitigate this effect, we attached a Bow Guide (HorACE,  BG-140) to the violin body to support the bow's trajectory. In order to have the participants concentrate more on the sound, we also asked to listen directly to the sound of the silent violin through earphones during the experiments.

\subsection{Results and Discussion}
To evaluate the system, we first used the measured data to evaluate two items: (\ref{p1}) whether the magnitude of bow pressure improved, and (\ref{p2}) whether the difference in bow pressure at each bow position decreased. We did not include the data of S1 in the comparison because we provided it to familiarize the participants with the use of the system and the instrument itself. Next, we evaluated system usability quantitatively using the System Usability Scale (SUS)~\cite{sus} and qualitatively through interviews.

\subsubsection{Magnitude of Bow Pressure}
\label{sec:mag}
We compared three conditions for the magnitude of bow pressure during measurement: baseline (S0), after practice (S2), and after day (S3). The Wilcoxon signed-rank test was conducted at a significance level of 5\%, with the Bonferroni correction applied for multiple comparisons of bow pressures between the three sessions. 

\begin{figure}[tb]
\centering
\includegraphics[width=0.75\columnwidth]{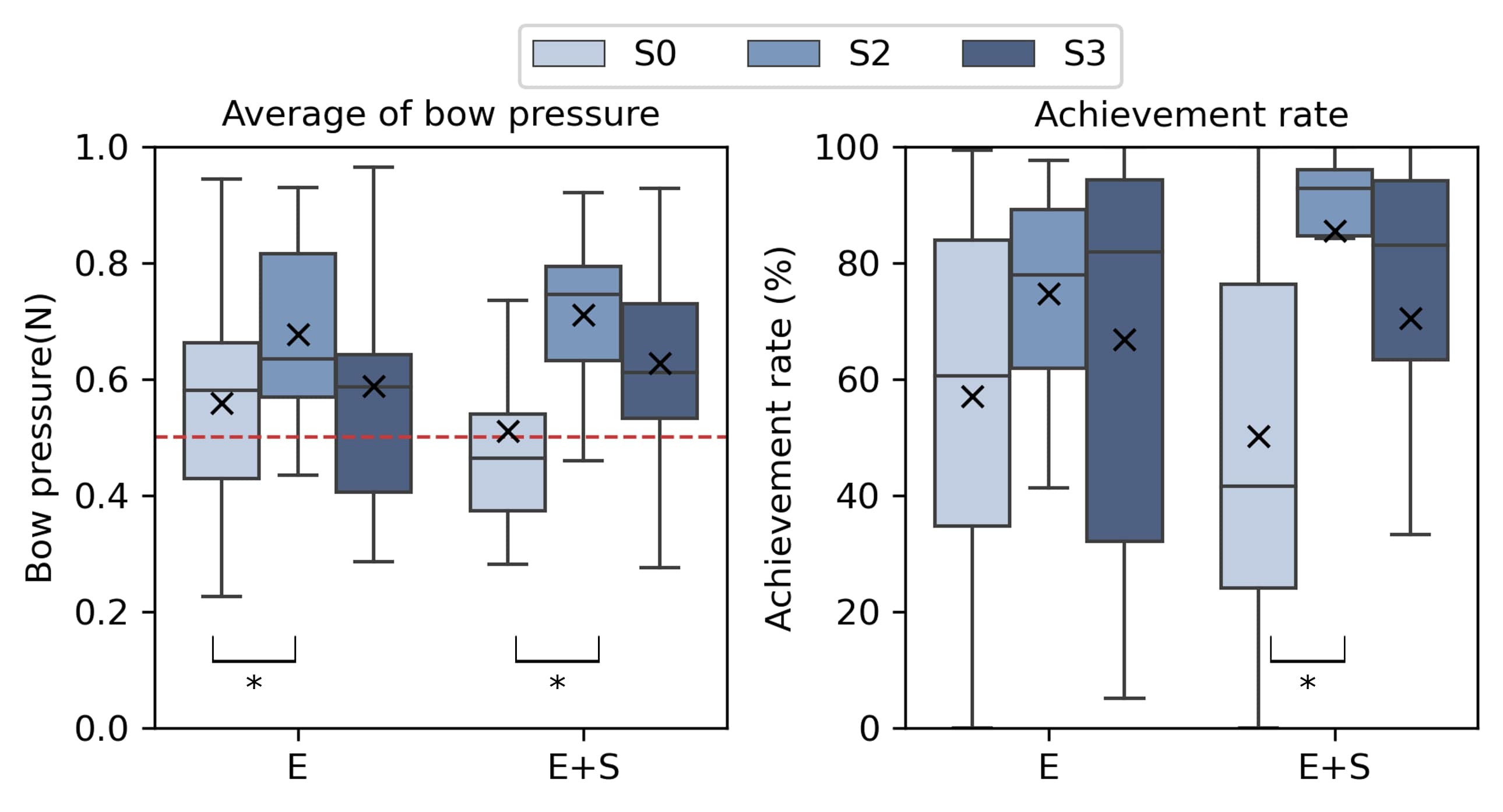}
\caption{The average bow pressure in each session (left) and percentage of bow pressure above threshold (right) (*: p$<$0.05)}
\label{fig:press}
\end{figure}

Figure~\ref{fig:press} shows the average bow pressure in each session and the percentage of bow pressure applied above the threshold during the measurement (achievement rate). The left figure shows the mean bow pressure of both groups was significantly greater after practice (S2) than before practice (S0). It suggested that after S0, they tried to increase the bow pressure because they had instructions to keep the bow pressure constant above a certain level and the relationship between pressure and sound. However, there was no significant difference between day 1 (S0, S2) and day 2 (S3), and the mean bow pressure in S3 for the E group returned to the same level as in S0. The average bow pressure of the E+S group was greater in S3 than in S0. The E+S group that practiced using the system would be more aware of bow pressure and tried to increase it even when the system was not in use. 

Next, in the figure on the right, the E+S group had a significantly larger percentage of bow pressure above the threshold after practice (S2) than before practice (S0). S2 was measured using the system, and the E+S group practiced while receiving feedback that the color of the graph changed to blue when the bow pressure fell below the threshold. This result indicates that the system was particularly effective in increasing the bow pressure above the threshold. The achievement rate in S3 was not significantly different from S0, but it was higher than S0, suggesting that the effect of the system was sustained a little without it. On the other hand, the E group did not know how high the pressure threshold was and did not receive any feedback during the practice, so there was no difference between sessions.

\subsubsection{Difference in Bow Pressure at Each Bow Position}
\label{sec:diff}
\begin{figure}[tb]
\centering
\includegraphics[width=0.45\columnwidth]{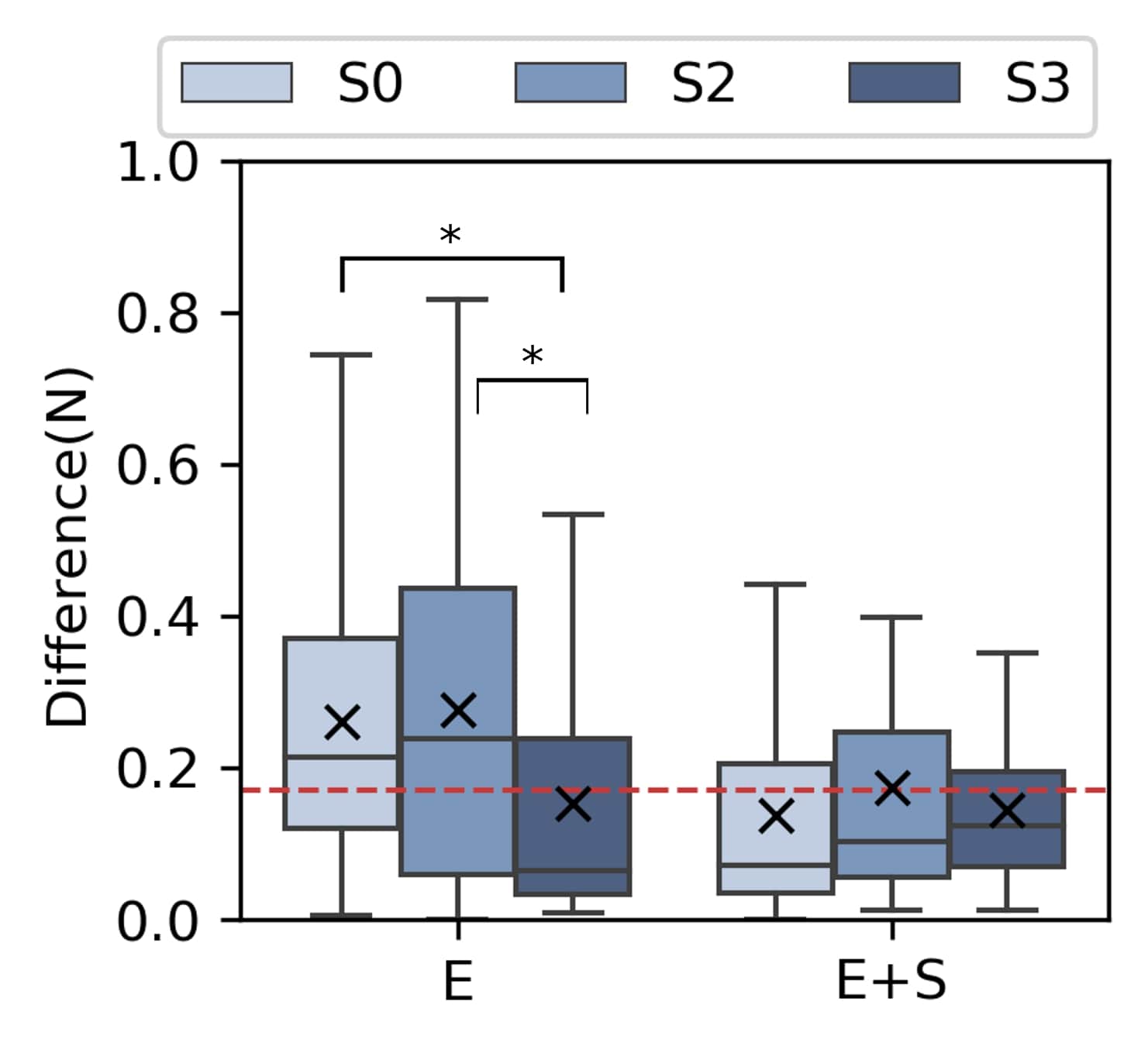}
\caption{Difference in average bow pressure for each bow position (*: p$<$0.05)}
\label{fig:diff}
\end{figure}

We compared three conditions for differences in bow pressure at each bow position: baseline (S0), after practice (S2), and after day (S3). We conducted the Wilcoxon signed-rank test at a significance level of 5\%, with the Bonferroni correction applied for multiple comparisons of bow pressures between the three sessions. Figure~\ref{fig:diff} shows the distribution of the largest difference in average bow pressure at each bow position. During the experiment, the system displayed the difference in bow pressure for each round trip, so we separated the data by one round trip and calculated the difference in bow pressure for each. The difference in bow pressure of the E group was significantly smaller on day 2 than on day 1. On the second day, the participants were more accustomed to holding and moving the instrument and applying bow pressure evenly regardless of bow position. On the other hand, there were no significant differences between any of the sessions in the E+S group; the difference in bow pressure at S0 in the E+S group was often already below the threshold of 0.17 $N$, so the system feedback from the system that changed the color of the graph was less frequent or it was not easy to reduce the difference any further. However, both groups showed the smallest variation in bow pressure difference at S3, suggesting that they were able to apply bow pressure evenly through practice.

Next, we examined the effect of feedback during the use of the system. During the measurement, the system updated a graph of the distribution of bow pressure at each position and the magnitude of the difference after each round trip. If the difference was greater than the threshold (0.17 $N$), the system provided feedback that the color of the graph changed to red. We calculated the probability (improvement rate) that the difference in bow pressure became smaller in the next round trip from when the feedback was given. We described the improvement rate as the ratio of times the bow pressure difference was greater than the threshold by the third round trip out of the four round trips to the number of times the bow pressure difference became smaller in the next round trip.

\begin{figure}[tb]
\centering
\includegraphics[width=0.7\columnwidth]{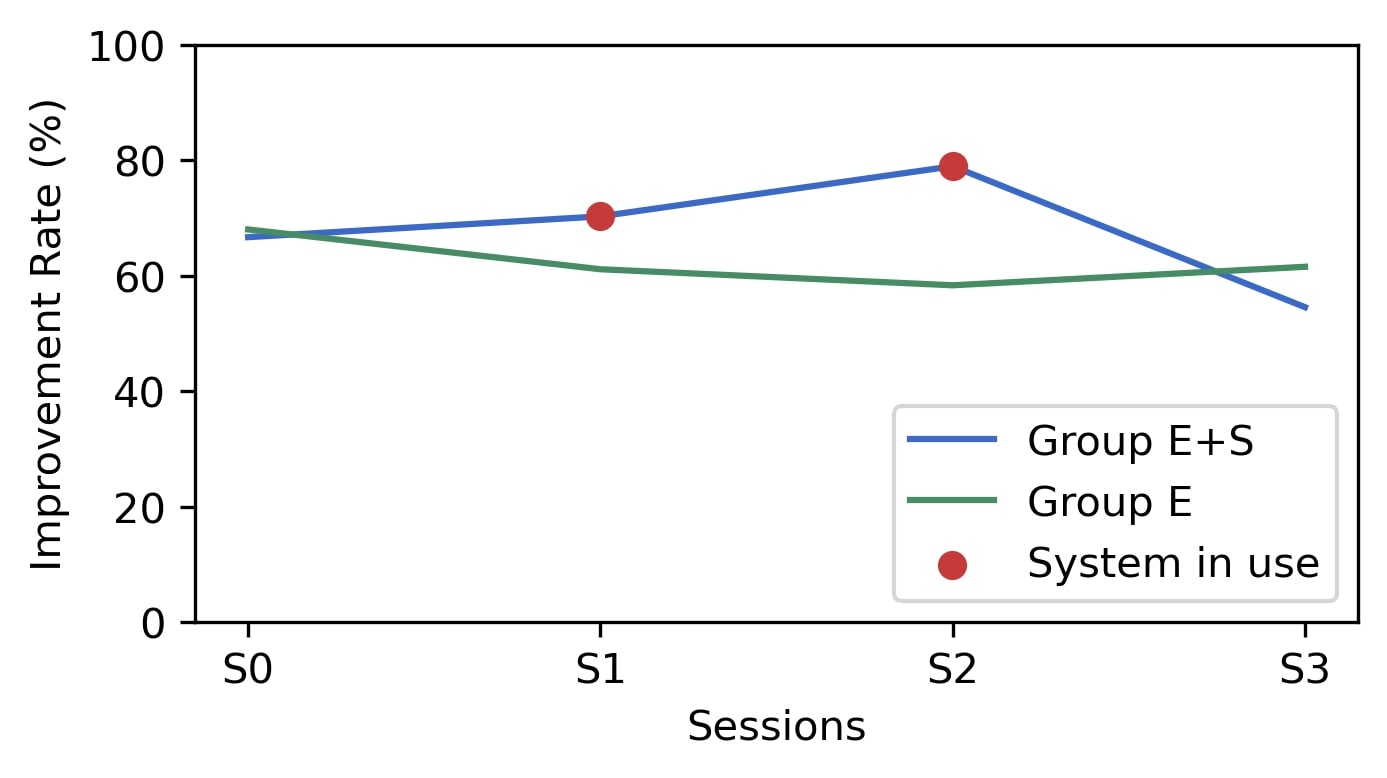}
\caption{Comparison of improvement rate with and without use of the system}
\label{fig:improvement}
\end{figure}

Figure~\ref{fig:improvement} shows the results of the improvement rates for each session. Results include data from S1 during practice when the E+S group used the system. The improvement rate increased while using the system and was greater than all the results of the E group. However, the E+S group showed a decrease in the improvement rate in S3 when the system was not in use. This result indicated that the user tried to reduce the bow pressure difference in the next round trip only when the system provided feedback changing the color of the graph in red.

\subsubsection{Sound Quality Evaluation of Recordings}
In Section~\ref{sec:mag} and~\ref{sec:diff}, we discussed whether the quantitative index of bow pressure we focused on improved. To verify whether the improvement in the quantitative values led to an improvement in performance as well, we experimented to evaluate the sound quality of the performance. We asked experienced players to listen to and evaluate the sound recorded during the S0, S2, and S3 measurements.

First, we selected the more stable sound source out of the two measurements in each session. We failed to record the sounds of one participant in the E group, so we excluded their data. Therefore, we collected 39 sound sources (three sessions $\times$ (six participants in E group + seven participants in E+S group)). We edited the sources first. We cut off the first few seconds of each recording and extracted the 30 seconds of the latter half, which was considered more stable.

We recruited 10 experienced violin players (three males and seven females, average age 23.8 $\pm$ 1.89 years). They had an average of 17.1 $\pm$ 3.51 years of violin experience, and all had more than 10 years of experience. We asked experienced players to listen to the recordings on the questionnaire form and evaluate their sound quality on a scale of 1--10. We asked them to rate whether the appropriate bow pressure was applied evenly, ignoring tempo deviations and parts of the recording that touched other strings. We specified no medium or volume level for listening to the recordings. We presented all sound recordings in random order.

\begin{figure}[tb]
\centering
\includegraphics[width=0.45\columnwidth]{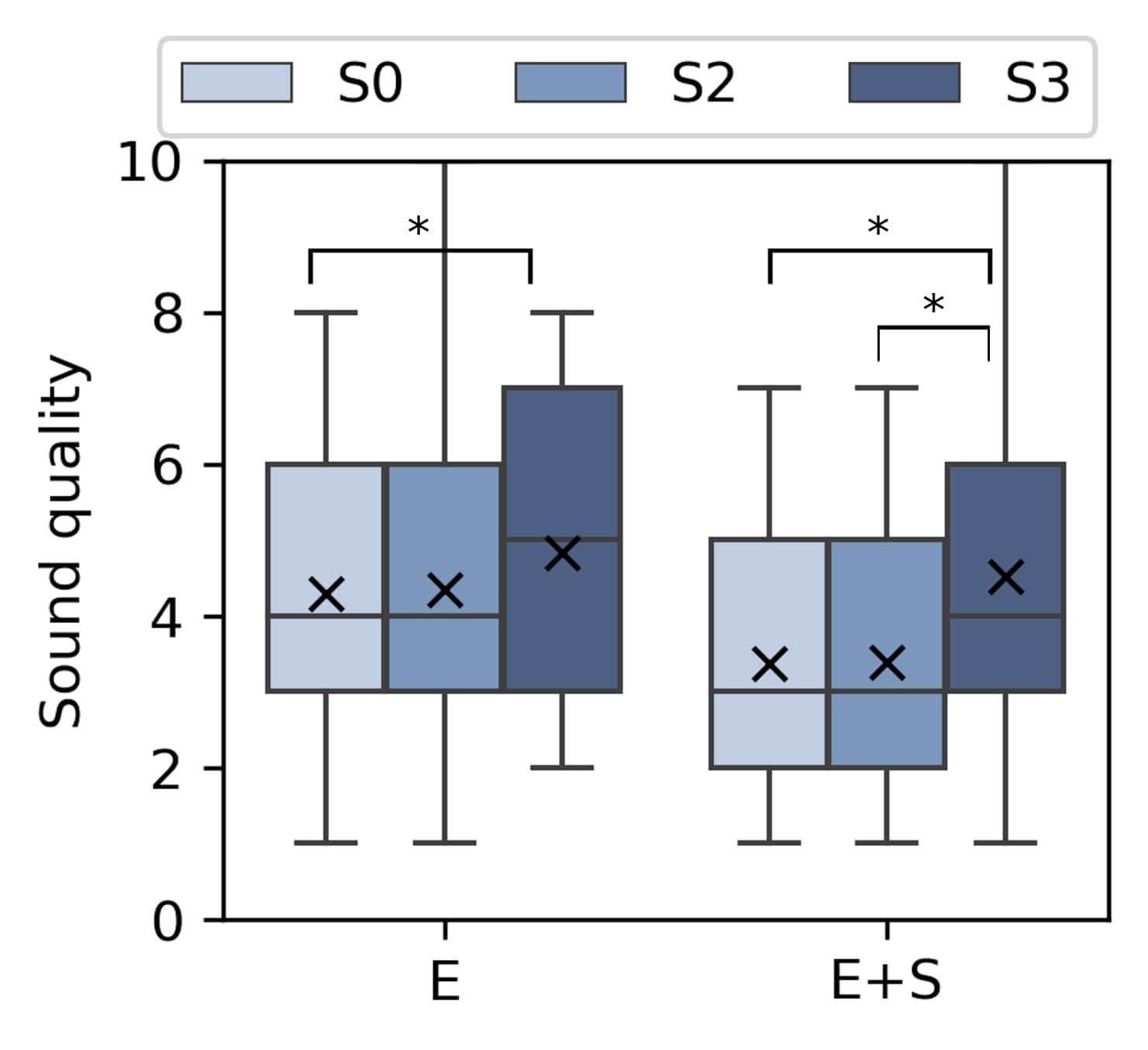}
\caption{Sound quality evaluation of recordings in each session (*: p$<$0.05)}
\label{fig:recordings}
\end{figure}

Figure~\ref{fig:recordings} shows the results. Higher values for sound quality ratings indicate better sound quality. We compared three conditions for the sound quality: baseline (S0), after practice (S2), and after day (S3). The Wilcoxon signed-rank test was conducted at a significance level of 5\%, with the Bonferroni correction applied for multiple comparisons of sound quality between the three sessions. 

The experienced players rated the sound quality of both groups significantly better on the second day of performance than before practice. We found that the practice allowed for a more even application of bow pressure in terms of sound quality. In the E+S group, there were also significant differences in S2 and S3, but the quality ratings of S0 and S2 remained almost the same. In a post-experiment interview, two participants in the E+S group indicated that they paid more attention to the sound when they were not using the system than when using it. Another three participants indicated that they only paid attention to the sound and tried to improve it when they touched other strings or heard a stuck sound. When they played while using the system, they paid less attention to the sound and did not improve the sound quality. In the Blanco et al. study~\cite{Blanco}, visual feedback of bow angle improved bowing motion but worsened sound quality. Practicing with a system that provides visual feedback may interfere with attention to sound.

\subsubsection{System Usability}
After S2, we asked the E+S group who used the system to answer a usability questionnaire to calculate the SUS score. The resulting calculated SUS score was 78.9. Since the score was above the average score of 68, the usability of the system was generally good.

Next, we summarize the post-experiment interviews. First, several participants commented that the system's UI was easy to understand because the color changes allowed them to judge whether it was good or bad. On the other hand, some did not know how to understand the central box-and-whisker chart in the first place, or how to improve it in concrete terms. Indeed, the box-and-whisker graph only presented data and did not provide color feedback, so users were burdensome because they had to make their judgments about how to adjust their bow pressure. Some commented that it would be rewarding to provide how much they had improved as a score, even if it was not specific instructions.

When asked which graph they looked at the most in the system, half of the participants indicated that they looked at the left graph of bow pressure and the other half looked at the box-and-whisker diagram in the center. The reasons for looking at the bow pressure graph were that the magnitude of bow pressure often did not reach the threshold and that the box-and-whisker diagram was not updated as frequently. The reasons for looking at the box-and-whisker diagram were that the user did not have to constantly check the left graph because the pressure achieved the threshold or it could be judged by its color. In this way, the participants viewed the different graphs viewed depending on their achievement status. Therefore, we thought we could reduce the burden on the user by changing the number, type, and size of the graphs displayed, depending on the user's achievement status and the items he or she wanted to achieve. There was an opinion about the center graph that it would be good to have a function to increase the number of divisions of the bow position to increase the level of difficulty according to the user's level.

Most participants found the violin itself more difficult than the system itself. Some commented that there was a physical burden because they could not get used to holding the instrument, and others said that there were so many things to take care of. At the beginning of the experiment, we explained how to hold the instrument and basic playing movements, but it was not enough to get them used to handling the instrument. In addition, many participants were distracted by factors other than bow pressure, such as bow trajectory and bow angle which led them to pay attention to moving the bow straight or playing one string without touching the other strings. Participants with no musical experience also commented that it was difficult to listen to the metronome and play along with it. In the experiment, the screen of the system was placed in the direction of the neck of the violin, so that the participants could see the part where the bow and strings made contact with each other and the screen without changing the direction of their faces. However, some users commented that it was difficult to shift their gaze to look at the screen and the instrument alternately and that it was difficult to adjust the playing motion while looking at the screen. From these opinions, the user of this system should be a beginner who is a little more familiar with violin playing. This system was too much of a burden for someone who had never played the violin at all, as was the case in this experiment.

Two participants also indicated that the system feedback did not match the sound quality. Specifically, when the user increased bow pressure as per the system feedback, the sound became dirty or did not sound good even when the pressure and difference achieved the threshold. Since the tone of the violin is not affected by bow pressure alone, we thought that by providing feedback on other bowing motion elements and scoring the tone, the participants could practice without being limited to bow pressure values alone.

\section{Limitations and Future Works}
We installed a load cell under the bridge to measure bow pressure. However, it is difficult to install such a sensor on a real violin. In addition, using this method, we can measure bow pressure only when playing an open string because the pressure changes when the player presses the strings down with their fingers. As described in Section~\ref{sec:relatedworks}, there are many research methods for measuring bow pressure that are not affected by fingering movements and that can be easily attached to existing violins~\cite{Young,Pardue}. By using these methods, it is possible to measure bow pressure and use systems with existing instruments while practicing fingering movements.

We used motion capture for bow position measurement. However, this method has high monetary and spatial costs and sometimes fails to measure due to sunlight and infrared radiation. There are other methods other than motion capture for bow position measurement, such as using optical sensors~\cite{Provenzale} or resistance wires~\cite{askenfelt1}. Combining these methods would make the bow position measurement easier.

In Section~\ref{sec:feedback}, we showed that teaching the difference between experienced and novice players was effective in improving the magnitude of bow pressure and the difference due to position. In the experiment, we gave both groups instructions regarding the results of Section~\ref{sec:begexp}. We need to compare the results with practice without instruction to confirm the effect of practice more clearly.

Next, in the implemented system, we set the threshold for giving feedback based on the measurement results of the experiment in Section~\ref{sec:begexp}. However, some participants achieved that threshold from the beginning, or their bow pressure was improved by other factors, such as bow trajectory support attachment or pre-practice instruction. The lack of significant differences might be due to the predefined threshold being inappropriate or some participants were not targeted users of the system. Therefore, the threshold settings should be varied according to the level of the user. In addition, the system should be implemented to change the UI according to the user's practice objectives and to change the level of difficulty according to the user's level, as mentioned in the interviews.

\section{Conclusion}
This study focused on violin bow pressure. First, we measured the bow pressure of experienced players and beginners and analyzed the differences by comparing them. The results showed that the experienced players consistently applied a higher minimum bow pressure and, particularly during down-bows, applied higher bow pressure only at the beginning of the stroke, then quickly reduced it while keeping the bow pressure steady until reaching the tip. In addition, the experienced players exhibited faster bow direction changes and maintained a consistent bow speed, regardless of the bow position. 

Second, we investigated the effect on the practice of teaching beginners the characteristics of experienced players that we had identified. In particular, we explained and taught the characteristics of maintaining bow pressure above a certain level and keeping bow pressure constant regardless of bow position using the system we developed. The results showed that the magnitude of bow pressure was significantly greater in the post-teaching sessions. In particular, the group that practiced with the system applied more bow pressure than the threshold value indicated in the system. This group also tended to have greater bow pressure than before practice when measured without the system later, although there was no significant difference. The difference in bow pressure for each bow position was significantly smaller on the second day compared to on the first day in the group that received only the explanation. The users tended to try to reduce the difference when there was feedback from the system. The sound quality also improved after the practice compared to before.

In the future, we will consider adding functions to the system that change the user interface or the difficulty level according to the user's level.

\section*{Acknowledgment}
Part of this work was supported by JST PRESTO (Grant Number JPMJPR2134).

\bibliographystyle{elsarticle-num} 
\bibliography{bib.bib}

\end{document}